\documentclass[twocolumn]{aastex62}

\usepackage{CJKutf8}
\newcommand{\dml}{dm-$\lambda$}
\newcommand{\cntext}[1]{\begin{CJK}{UTF8}{gbsn}#1\end{CJK}\kern-1ex}

\graphicspath{{./}{figures/}}
\usepackage[normalem]{ulem}

\received{June 27, 2019}
\revised{August 12, 2019}
\accepted{August 15, 2019}
\submitjournal{ApJ}

\shorttitle{Solar Flare Termination Shock: Split-Band Feature}
\shortauthors{Chen et al.}

\turnoffedit

\begin{document}

\title{Radio Spectroscopic Imaging of a Solar Flare Termination Shock:
Split-Band Feature as Evidence for Shock Compression}

\correspondingauthor{Bin Chen}
\email{bin.chen@njit.edu}

\author[0000-0002-0660-3350]{Bin Chen (\cntext{陈彬})}
\affiliation{Center for Solar-Terrestrial Research, New Jersey Institute of Technology, 323 M L King Jr Blvd, Newark, NJ 07102-1982, USA}

\author[0000-0002-9258-4490]{Chengcai Shen (\cntext{沈呈彩})}
\affiliation{Harvard-Smithsonian Center for Astrophysics, 60 Garden St, Cambridge, MA 02138, USA}

\author[0000-0002-6903-6832]{Katharine K. Reeves}
\affiliation{Harvard-Smithsonian Center for Astrophysics, 60 Garden St, Cambridge, MA 02138, USA}

\author[0000-0003-4315-3755]{Fan Guo (\cntext{郭帆})}
\affiliation{Los Alamos National Laboratory, P.O. Box 1663, Los Alamos, NM 87545, USA}
\affiliation{New Mexico Consortium, Los Alamos, NM 87544,  USA}

\author[0000-0003-2872-2614]{Sijie Yu (\cntext{余思捷})}
\affiliation{Center for Solar-Terrestrial Research, New Jersey Institute of Technology, 323 M L King Jr Blvd, Newark, NJ 07102-1982, USA}

\begin{abstract}
Solar flare termination shocks have been suggested as one of the promising drivers for particle acceleration in solar flares, yet observational evidence remains rare. By utilizing radio dynamic spectroscopic imaging of decimetric stochastic spike bursts in an eruptive flare, Chen et al. found that the bursts form a dynamic surface-like feature located at the ending points of fast plasma downflows above the looptop, interpreted as a flare termination shock. One piece of observational evidence that strongly supports the termination shock interpretation is the occasional split of the emission band into two finer lanes in frequency, similar to the split-band feature seen in fast-coronal-shock-driven type II radio bursts. Here, we perform spatially, spectrally, and temporally resolved analysis of the split-band feature of the flare termination shock event. We find that the ensemble of the radio centroids from the two split-band lanes each outlines a nearly co-spatial surface. The high-frequency lane is located slightly below its low-frequency counterpart by $\sim$0.8 Mm, which strongly supports the shock-upstream--downstream interpretation. Under this scenario, the density compression ratio across the shock front can be inferred from the frequency split, which implies a shock with a Mach number of up to 2.0. Further, the spatiotemporal evolution of the density compression along the shock front agrees favorably with results from magnetohydrodynamics simulations. We conclude that the detailed variations of the shock compression ratio may be due to the impact of dynamic plasma structures in the reconnection outflows, which results in distortion of the shock front. 

\end{abstract}


\section{Introduction} \label{sec:intro}

In a solar flare, a fast-mode shock can form when reconnection outflows, as a result of fast magnetic reconnection, impinge upon the top of flare arcades, provided that the outflow speed exceeds the local fast-mode magnetosonic speed. In a steady-state picture, the shock is perceived to be a standing shock above the looptop and is usually referred to as a flare ``termination shock.'' Flare termination shocks were long predicted in numerical simulations of flares \citep{1986ApJ...305..553F,1988SoPh..117...97F,1986ApJ...302L..67F,2011PhPl...18i2902W,2015ApJ...805..135T,2016ApJ...823..150T,2017ApJ...848..102T,2018ApJ...869..116S} and were frequently invoked in some of the most well-known schematics of the standard flare model \citep[e.g.,][]{1994Natur.371..495M,1995ApJ...451L..83S, 1996ApJ...466.1054M,2000JGR...105.2375L}. 
They have also been suggested as an outstanding candidate for driving particle acceleration \citep{1986ApJ...305..553F,1995ApJ...451L..83S,1997ApJ...485..859S,1998ApJ...495L..67T,2009A&A...494..669M,2009A&A...494..677W,2012ApJ...753...28G,2013ApJ...769...22L,2013PhRvL.110e1101N,2013ApJ...765..147P} and plasma heating \citep{1994Natur.371..495M,2015ApJ...800...54G}. 

However, observational evidence of flare termination shocks remains rare. This dearth of observations is because, first of all, the termination shocks are confined at the front of highly collimated reconnection outflows. Hence, the termination shocks are expected to be small in its spatial extension. In recent two-dimensional (2D) magnetohydrodynamics (MHD) simulations, the size of termination shocks is found to be only a small fraction of that of the flare arcades when the current sheet is viewed edge on ($\lesssim$20\%; see, e.g., \citealt{2015Sci...350.1238C,2015ApJ...805..135T,2016ApJ...823..150T,2018ApJ...869..116S}), and the thickness of the shock front itself would be well below the resolution of current instruments. Second, their thermal signatures are difficult to distinguish from the highly dynamic and complex environment in the looptop region, where vigorous energy conversion and momentum transfer are expected to occur \citep[e.g.,][]{2016ApJ...823..150T,2019ApJ...875...33H}. 

Nonthermal HXR sources and bright thermal extreme ultraviolet (EUV)/X-Ray sources at or above the flare looptops \citep[e.g.,][]{1994Natur.371..495M,2013ApJ...767..168L,2015ApJ...800...54G} have been frequently considered as possible signatures for flare termination shocks, as they are excellent drivers for both intense plasma heating and electron acceleration. In fact, in the celebrated work by \citet{1994Natur.371..495M}, who first reported a coronal HXR source located well above the soft X-ray (SXR) flare arcades during the impulsive phase of a flare (``above-the-looptop HXR source'' hereafter), the authors interpreted the observed HXR source as the signature of a flare termination shock. There were also reports of the ``superhot'' ($>$30 MK) X-ray sources located above the looptop, which, as argued by \citet{2010ApJ...725L.161C} and \citet{2014ApJ...781...43C}, are likely heated directly in the corona by plasma compression or shock (although the termination shock is not specifically mentioned). In another work \citep{2015ApJ...800...54G}, by constructing differential emission measure maps of a ``candle-flame'' shaped post-flare arcades using multiband imaging data from the Atmospheric Imaging Assembly (AIA) aboard the \textit{Solar Dynamics Observatory (SDO}; \citealt{2012SoPh..275...17L}), the authors reported a localized excess of pressure and density at the flare looptop, as well as a long-lasting pressure imbalance between either half of the flare arcade. Both signatures are indications of a long-duration standing shock at the looptop, and a fast-mode flare termination shock was suggested as a favorable possibility. However, it remains difficult to associate these EUV/HXR sources with flare termination shocks due to the lack of more distinctive shock signatures.

One approach to probe the termination shocks is to use high-resolution UV/EUV imaging spectroscopic observations. This method is based on the expectation that plasma flows at the upstream and/or downstream side of the shock should exhibit profound signatures in the observed Doppler speeds. In order to make such measurements, however, the slit of the (E)UV imaging spectrograph needs to be extremely fortuitously placed at the close vicinity of the termination shock with an orientation preferably across the shock surface at the right moment. Such observations are very rare: to our knowledge, there have been only a few reports of this kind that possibly support a termination shock interpretation \citep{2011ApJ...741..107H,2013ApJ...776L..11I,2014ApJ...797L..14T,2018ApJ...865..161P}. Of particular interest is the detection of high-speed ($>$100~km~s$^{-1}$) redshift signatures in the \ion{Fe}{21} 1354.08 \AA\ line emitted by hot flaring plasma ($\sim$10 MK) at the (above-the-)looptop region \citep{2013ApJ...776L..11I,2014ApJ...797L..14T,2018ApJ...865..161P}. As argued by \citet{2018ApJ...865..161P}, they may be associated with the heated plasma in the shock downstream. This argument is also supported by \citet{2017ApJ...846L..12G}, who used an MHD model to study the manifestation of the termination shock in the synthetic \ion{Fe}{21} line profiles. 

Another excellent probe for shocks lies in coherent radio bursts. Nonthermal electrons in the vicinity of a shock, either accelerated locally by the shock or transported from elsewhere, can be unstable to the production of Langmuir waves, providing that the electrons have an anisotropic distribution favorable for wave growth. The Langmuir waves can subsequently convert into transverse electromagnetic waves via a variety of nonlinear wave--wave conversion processes, manifesting as bright radio bursts at frequencies near the local plasma frequency $\nu_{pe}=(e^2n_e/\pi m_e)^{1/2}\approx 8980\sqrt{n_e}$ Hz or its harmonic. The most well-known radio bursts of this type are the type II radio bursts, which are associated with coronal shocks driven by fast coronal mass ejections (CMEs) or flare-related blast waves (e.g., \citealt{1950AuSRA...3..387W,1983ApJ...267..837H,1984A&A...134..222G,1985srph.book.....M,1999GeoRL..26.1573B,2001A&A...377..321V,2006A&A...448..739V, 2002A&A...384.1098C,2011A&A...531A..31N,2012ApJ...750...44B,2012A&A...547A...6Z,2018A&A...615A..89Z,2019NatAs...3..452M}; see also reviews by \citealt{1995LNP...444..183M,2003SSRv..107...27C}). In the radio dynamic spectrum, the type II radio bursts appear as a bright drifting structure toward lower frequencies in time (i.e., $d\nu/dt<0$) as the shock propagates into the higher corona with a decreasing plasma density. 

\citet{2002A&A...384..273A} first reported type-II-burst-like structures with little or no frequency drift, and interpreted them as radio emission associated with a ``standing'' shock, likely a flare termination shock at the top of flare arcades. Three additional observations of this kind have been subsequently reported \citep{2004ApJ...615..526A,2009A&A...494..669M, 2009A&A...494..677W}. Radio imaging of the bursts at meter wavelengths by the Nan\c{c}ay Radioheliograph (NRH; \citealt{1997LNP...483..192K}) placed the burst sources somewhere in the corona above the flaring loops, yet these studies were limited by imaging at a single frequency with an angular resolution (100$''$--200$''$) at least one order of magnitude larger than the expected size of the termination shocks.

\begin{figure}[ht!]
\epsscale{1.1}
\plotone{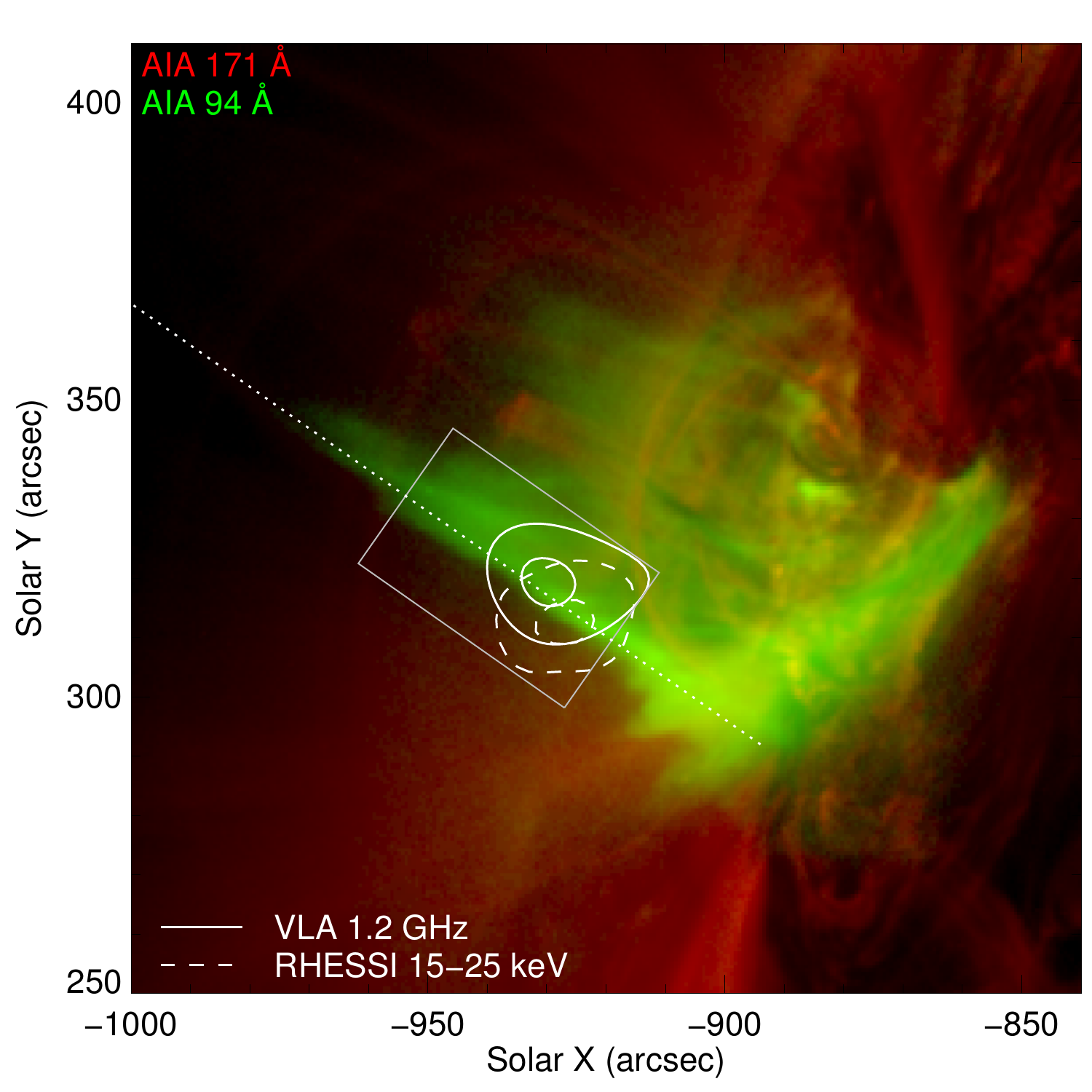}
\caption{Overview of the eruptive C1.9 flare event on 2012 March 3. Background shows \textit{SDO}/AIA 171 \AA\ (red) and 94 \AA\ (green) images at 18:30 UT. The solid and dashed contours are a radio stochastic spike burst source observed by VLA at 1.2 GHz (88\% and 90\% of the maximum) and \textit{RHESSI} 15--25 KeV HXR source (60\% and 90\% of the maximum), respectively. The rectangle shows the field of view of Figure \ref{fig:schem}(B). (Adapted from \citealt{2015Sci...350.1238C}.)\label{fig:overview}}
\end{figure}

\begin{figure*}[ht!]
\epsscale{1.1}
\plotone{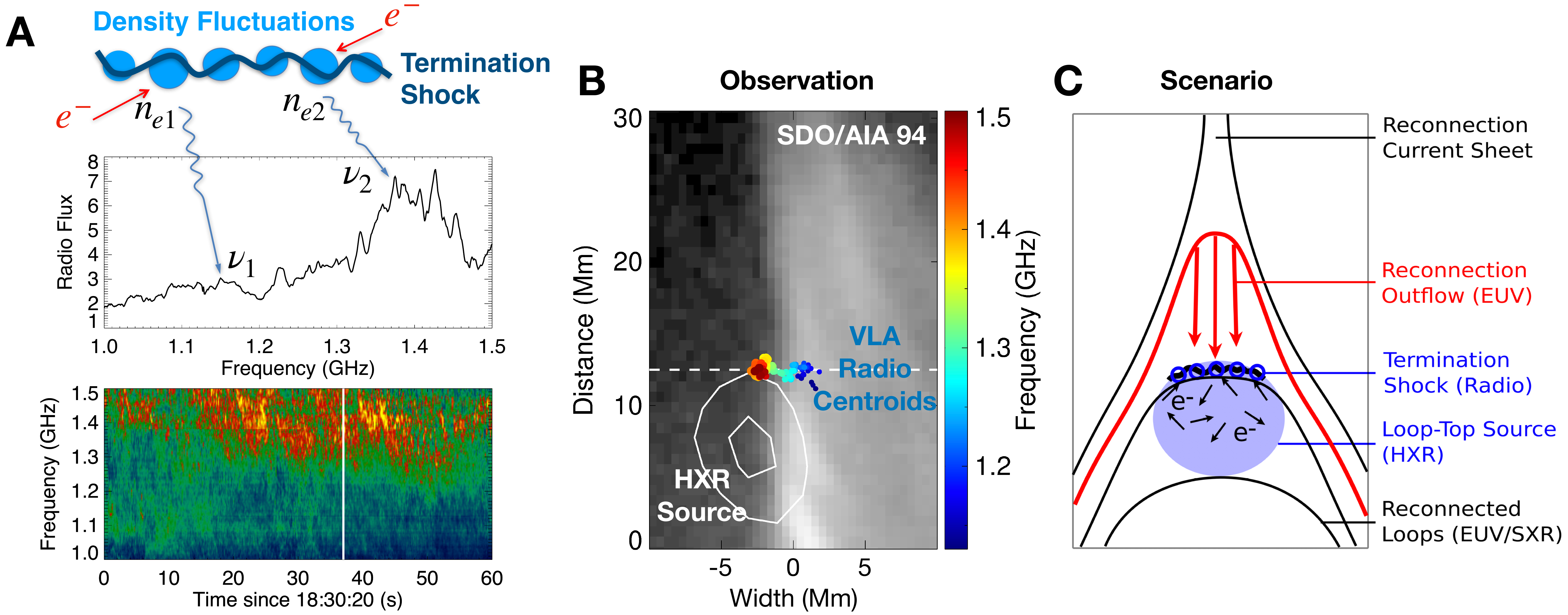}
\caption{Schematic of the formation of the \dml\ stochastic spike bursts in the close vicinity of the flare termination shock. (A) At each time, small-scale density fluctuations on the shock surface emit radio bursts at different frequencies due to plasma radiation. The instantaneous radio spectrum is obtained at a time indicated by the vertical white line in the radio dynamic spectrum. (B) The instantaneous distribution of the spike source centroids at a given time (18:30:57 UT; same as the time in (A)) delineate the shock surface above a coronal HXR source. The field of view is indicated by the rectangle in Figure \ref{fig:overview} rotated to an upright orientation, with the vertical axis aligning along the direction of the dotted line in Figure \ref{fig:overview}. The horizontal and vertical axes (described later in Section \ref{sec:nature}) designated, respectively, as the $x$- and $y$-axes, form the coordinate system adopted for the rest of the paper. (C) A schematic of the termination shock in the context of flare reconnection and the associated emission signatures. (Adapted from \citealt{2015Sci...350.1238C}.) \label{fig:schem}}
\end{figure*}

The most decisive evidence of the presence of a flare termination shock and its possible role as an electron accelerator to date was provided by \citet{2015Sci...350.1238C}, who used the Karl G. Jansky Very Large Array (VLA; \citealt{2011ApJ...739L...1P}) to observe a slow-drift decimetric burst event occurred during the extended impulsive phase of a long-duration eruptive C1.9 flare (Figure \ref{fig:overview}). Thanks to VLA's capability of performing dynamic spectroscopic imaging at high angular resolution (10$''$--30$''$) and spectrometer-like spectral resolution (0.05--0.1\% at 1--2 GHz), combined with extremely high temporal cadence (50 millisecond at the time of observation), they were able to resolve the myriad sub-structures of the burst group as stochastic spike bursts (see bottom panel of Figure \ref{fig:schem}(A)), and pinpoint the source centroid of each spike burst with very high positional accuracy ($\lesssim$1$''$). The distribution of the spike burst centroids at any given time forms a dynamic surface-like feature located at the ending points of fast plasma downflows (with an average speed of $\sim$550 km s$^{-1}$ and maximum speed up to 850 km s$^{-1}$) and slightly above a coronal HXR source over the looptop (Figures \ref{fig:schem}(B) and (C)). The centroid of each spike burst at a central frequency of $\nu$ with a bandwidth of $\delta\nu$ is interpreted as plasma radiation due to linear mode conversion of Langmuir waves on a small-scale density structure with a mean density of $n_e$ and a fluctuation of $\delta n_e/n_e\approx 2\delta\nu/\nu$. The instantaneous distribution of the observed density fluctuation structures hence delineates the shock surface (see schematic in Figure \ref{fig:schem}(A)). More interestingly, a temporary disruption of the termination shock surface by the arrival of a fast plasma downflow coincides with the reduction of the radio and HXR flux for both the looptop and loop-leg sources, which strongly supports the role of the termination shock in accelerating electrons to at least tens of keV.

Here, we present a detailed study of the split-band feature of the same termination shock event reported by \citet{2015Sci...350.1238C}. This phenomenon is well-known for type II radio bursts: in the radio dynamic spectrum, the bursts sometimes split into two (occasionally more) finer, almost parallel lanes  \citep{1985srph.book.....M,2001A&A...377..321V,2002A&A...396..673V,2009ApJ...691L.151L,2012A&A...547A...6Z,2015ApJ...812...52D,2016ApJ...832...59K,2018ApJ...868...79C,2018A&A...615A..89Z}. One of the leading interpretations attributes the splitting high- and low-frequency branch (``HF'' and ``LF'' hereafter) to plasma radiation from the shock downstream and upstream region, respectively, because the downstream region of the shock has a higher plasma density (and higher plasma frequency since $\nu\propto\sqrt{n_e}$) due to shock compression \citep{1974IAUS...57..389S,1975ApL....16...23S}. VLA's dynamic spectroscopic imaging capability allows us to map, for the first time, the shock geometry associated with both the HF and LF split-band features simultaneously and investigate the shock compression in unprecedented detail.

The paper is structured as follows. After a brief overview of the event in Section \ref{sec:overview}, we present spectroscopic imaging observations from the VLA and the associated data analysis in Section \ref{sec:sbd}. In Section \ref{sec:nature}, we discuss the observational evidence that supports the interpretation of the split-band feature in terms of the shock upstream--downstream scenario. We then derive the spatially and temporally resolved shock compression ratio in Section \ref{sec:compression} and compare the results with MHD simulations. We discuss the lack of shock density compression signatures in EUV wavelengths in Section \ref{sec:euv}. Finally, we briefly summarize in Section \ref{sec:conclusion}.

\section{Observations and Interpretation}\label{sec:obs}
\subsection{Event Overview}\label{sec:overview}
The stochastic spike bursts event under study was recorded by the VLA between $\sim$18:20 and 18:40 UT during the extended impulsive phase of a C1.9-class flare on 2012 March 3. \citet{2014ApJ...794..149C} discussed the flare event and the associated magnetic flux rope eruption, and \citet{2015Sci...350.1238C} presented the first results of the dynamic termination shock using VLA and other multiwavelength data as well as MHD simulations. We refer readers to these papers for more details on this termination shock event as well as the related descriptions of the instrumentation and data analysis techniques. Here, we only provide a brief overview for the flare context. 

The flare originated in AR 11429 near the east limb, with the soft X-ray (SXR) peak occurring at $\sim$19:33 UT. It had an extended, $>$1 hr long impulsive phase, featured by multiple microwave bursts and long-duration HXR emission (above 12 keV). A variety of types of decimetric bursts, including the spike bursts, were recorded by the VLA in 1--2 GHz. The event was associated with the eruption of a ``hot channel'' \citep{2012NatCo...3E.747Z,2013ApJ...763...43C,2014ApJ...789L..35C} in the low corona seen by multiple \textit{SDO/AIA} passbands, with a hot envelope and a cool core, and a fast white light CME \citep{2014ApJ...794..149C}. As shown by \citet{2015Sci...350.1238C}, spectroscopic imaging of the decimetric stochastic spike bursts source at the looptop at any given time outlines a dynamic surface feature located slightly above the looptop HXR source, which coincides with the expected location of a flare termination shock in MHD modeling results (see Figure \ref{fig:schem}(B) and (C)). 

\subsection{Spectroscopic Imaging of the Split-band Feature}\label{sec:sbd}
\begin{figure}[ht!]
\epsscale{1.2}
\plotone{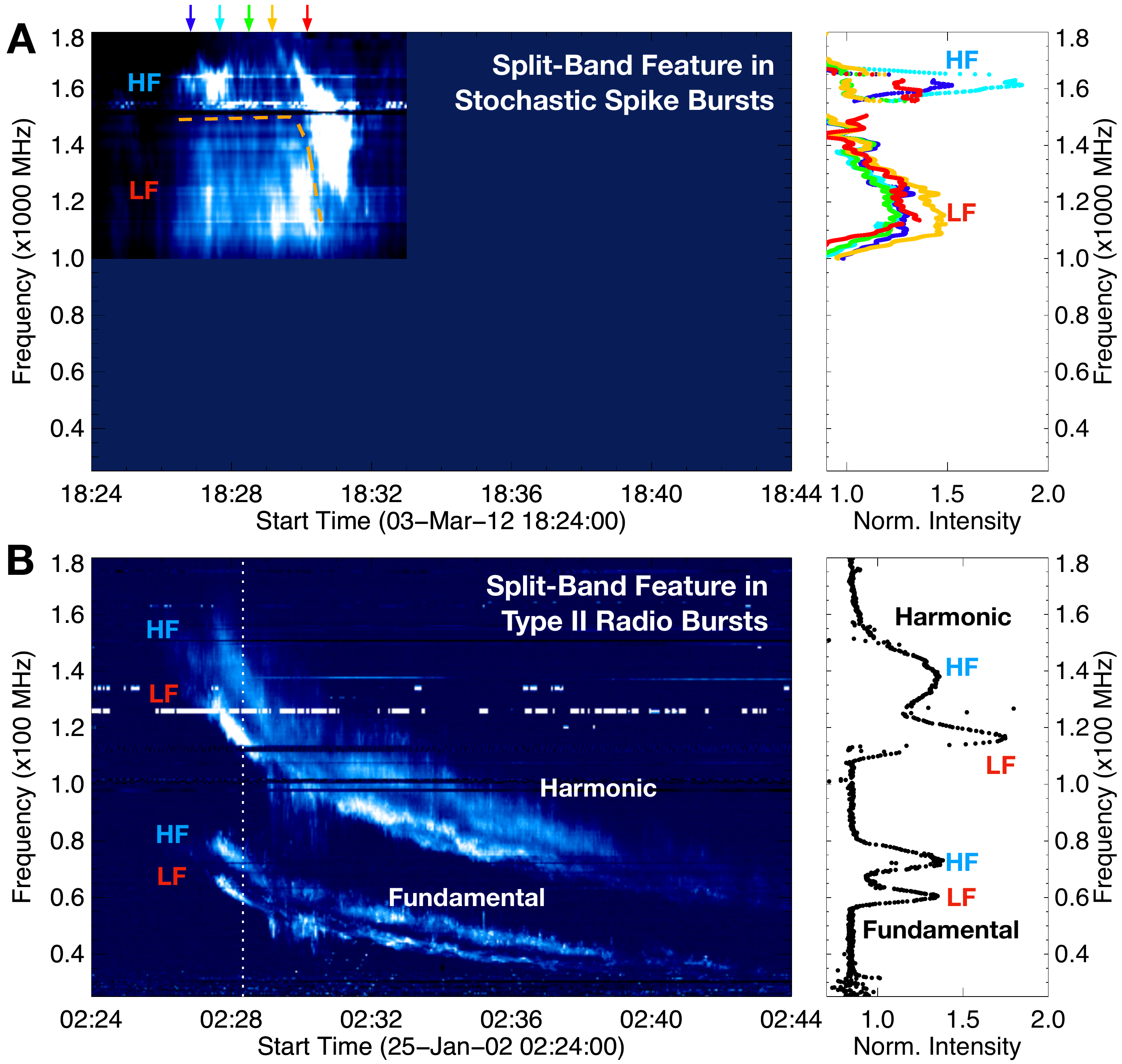}
\caption{Comparison between the slow-moving, flare-termination-shock-associated stochastic spike bursts (A) and a typical coronal-shock-associated type II radio burst with split-band features (B). The latter has both fundamental and harmonic plasma radiation signatures. Both dynamic spectra are shown with the same duration (20 minutes) and relative frequency range. Right panels show normalized intensity profiles as a function of frequency at selected times (arrows in (A) and dotted vertical line in (B)). The split-band features in both events manifest themselves as bright emissions separated in the frequency domain. The high- and low-frequency branches are labeled as ``HF'' and ``LF,'' respectively. A more detailed view of the split-band feature of the spike bursts is available in Figure \ref{fig:sbd_surface}(A). Note some bright horizontal features are radio frequency interference. \label{fig:ts_vs_tp2}}
\end{figure}

\begin{figure*}[ht!]
\plotone{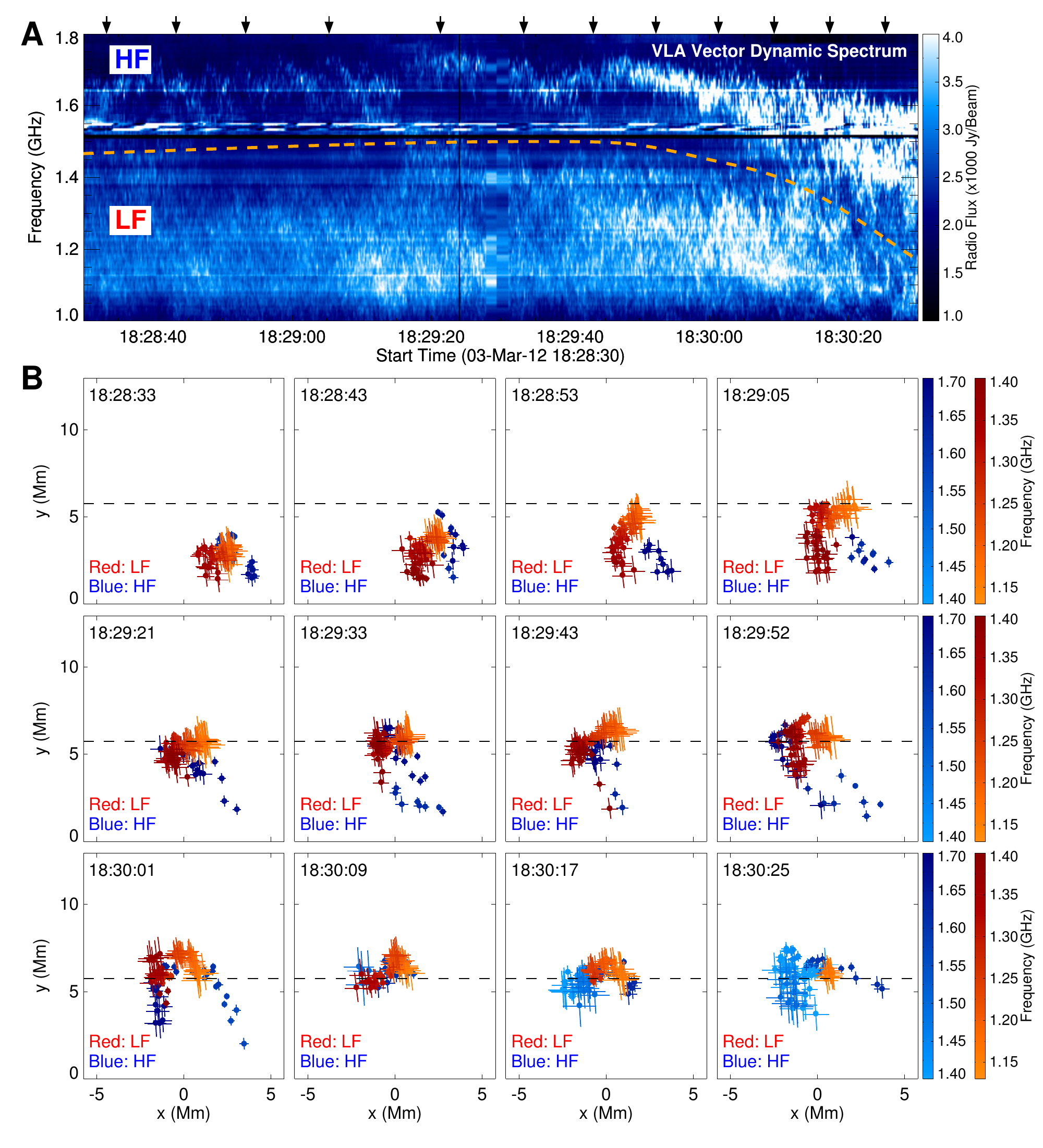}
\caption{Morphology and dynamics of both the high-frequency and low-frequency split-band features. (A) A more detailed view of the split-band feature in the vector dynamic spectrum. (B) Time sequence of the evolving surfaces associated with the HF and LF split-band features delineated by the centroids of each stochastic spike bursts. They are shown, respectively, in blue and red, with the corresponding frequency increasing from light to dark. Error bars show the positional uncertainties along the directions of the major and minor axes of the synthesized beam associated with each spike centroid. The reference horizontal line in each panel is at the same height as that shown in Figure \ref{fig:schem}(B). \label{fig:sbd_surface}}
\end{figure*}

By constructing the spatially resolved (or ``vector'') dynamic spectrum of the looptop radio source (see Supplementary Materials in \citealt{2015Sci...350.1238C} for details), the spectrotemporal feature intrinsic to the source is revealed: it consists of thousands of individual spike bursts, each has a duration of $<$50 ms and a small frequency bandwidth ($\delta\nu/\nu\approx1$--5\%). The group of spike bursts displays a slow overall drift in the dynamic spectrum, and moreover, appears to split into two finer lanes (Figure \ref{fig:sbd_surface}(A)). In order to put the observed spectrotemporal features of these slow-moving-termination-shock-produced bursts into the context of their counterparts associated with fast-propagating coronal shocks (i.e., type II radio bursts), in Figure \ref{fig:ts_vs_tp2}, we place the vector dynamic spectrum of our termination shock event side-by-side with a typical metric type II radio burst that also shows a clear split-band feature (recorded on 2002 January 25 by the Radio Solar Telescope Network in 25--180 MHz, which is one of the events studied by \citealt{2015ApJ...812...52D}). Both dynamic spectra are normalized to the same relative frequency range and are shown in the same time window (20 minutes). It appears that the termination-shock-associated radio bursts share many similarities in their appearance as the type II radio bursts, albeit they show a much slower overall frequency drift and relatively less-defined split-band lanes than their fast-propagating coronal counterparts. The slow overall frequency drift is related to the slow spatial movement of the termination shock at the looptop, and the less-defined split-bands may be attributed to the highly dynamic and likely turbulent plasma environment in the looptop region. 

VLA's high-cadence spectroscopic imaging capability provides the first opportunity to map \textit{both} the HF and LF lanes of the flare termination shock simultaneously. We first separate the two lanes in the frequency-time space based on their appearance in the vector dynamic spectrum (dashed line in Figure \ref{fig:sbd_surface}(A)). Following the technique described in \citet{2015Sci...350.1238C} for pinpointing the spike source centroids at different frequencies, at any given time, the distribution of the spike bursts at the HF and LF lane each delineates a coherent structure, shown in Figure \ref{fig:sbd_surface}(B) as blue- and red-color symbols, respectively. For each source centroid, the associated positional uncertainties along the direction of the major and minor axes of the synthesized beam are shown as the error bars. The uncertainties are estimated based on the relation $\sigma\approx\theta_{\rm FWHM}/({\rm S/N}\sqrt{8\ln 2})$, where $\theta_{\rm FWHM}$\ is the full width half maximum of the synthesized beam and S/N is the ratio of the peak flux to the root mean square noise of the image \citep{1988ApJ...330..809R,1997PASP..109..166C,2015Sci...350.1238C,2018ApJ...866...62C}. In this figure, we have excluded the centroids with uncertainties greater than 1$''$.3 ($\sim$1 Mm; or $\sim$20\% of the width of the shock surface), and removed those located at the edges of the spectral windows (which are subject to greater calibration errors as they have low instrumental response) as well as those affected by strong radio frequency interference (RFI). 

The HF and LF sources are nearly co-spatial with each other, both of which display a dynamic surface-like structure. In particular, the HF source develops from initially occupying only a small spatial section of the termination shock to ultimately dominating the emission at later times, while the LF source gradually diminishes as time progresses. The latter is probably due to the LF source drifting out of the lower-frequency boundary of our observing frequency range (1--2 GHz).

\subsection{Nature of the Split-band Feature} \label{sec:nature}
Our observations provide the \textit{first} picture of detailed instantaneous morphology for both the HF and LF split-band features associated with a flare termination shock. This observation is, to our knowledge, also the most detailed one to date of all reported split-band features in literature. In this subsection we discuss the physical nature of the split-band feature in the context of the flare termination shock, taking advantage of theories and models developed from numerous studies on coronal-shock-driven type II radio bursts. 

One leading interpretation for the split-band feature attributes the splitting HF and LF lanes to plasma radiation from the shock downstream and upstream regions, respectively, because the downstream region of the shock has a higher plasma density (and higher plasma frequency as $\nu\propto\sqrt{n_e}$) due to shock compression (\citealt{1974IAUS...57..389S,1975ApL....16...23S}; henceforth Scenario 1). Under this scenario, one could infer the shock density compression ratio $X$ from the observed frequency ratio of the HF and LF lanes via $X=n_2/n_1=(\nu_{\rm HF}/\nu_{\rm LF})^2=R_\nu^2$ (where $n_2$ and $n_1$ are density of the shock downstream and upstream, respectively), and in turn, estimate the shock Mach number based on the Rankine-Hugoniot jump conditions (see, e.g., Chapter 5 in \citealt{2014masu.book.....P}). 

Another popular scenario, which was initially proposed by \citet{1967PASAu...1...47M}, suggests that different portions of a large-scale shock front encounter the coronal environment with different physical properties, which may include plasma density, magnetic field, and shock geometry (henceforth Scenario 2). Consequently, observations of a type II burst with two or more finer bands in the dynamic spectrum can be expected if multiple portions of the shock front emit radio waves simultaneously. Later, \citet{1983ApJ...267..837H} interpreted the split-band features of type II bursts under the framework of the shock drift acceleration. In their picture, the radio bursts are produced by suprathermal electrons escaping along magnetic field lines upstream of the shock front. A split-band feature can be observed if there is a sufficient spatial separation between the type-II-emitting sources in the direction of the density gradient. More recently, using a numerical experiment, \citet{2005JGRA..110.1101K} reproduced type II bursts that had fine structures mimicking the split-band features. In their model, the split-bands were produced when a propagating shock front interacted with dense coronal loops, which effectively shifted the type II emission to higher frequencies at the site of interaction, hence leaving a void in the dynamic spectrum to resemble the split-band feature.

A key observational property to distinguish the two scenarios lies in the relative spatial locations of the HF and LF sources. In Scenario 1, the radio sources of the HF and LF branch must be located at the same section of the shock front, with a possibly small spatial separation between the two sources at the direction \textit{perpendicular to} the shock front. For Scenario 2, the HF and LF sources originate from different sections of the shock front, such that a spatial separation may be expected \textit{along} the shock front.

In the literature, using imaging observations, the spatial separation between the HF and LF split-band sources of type II radio bursts have been investigated \citep[e.g.,][]{2002A&A...383.1018K,2012A&A...547A...6Z, 2014ApJ...795...68Z,2018A&A...615A..89Z,2018ApJ...868...79C}. However, the lack of simultaneous imaging of the HF and LF sources at multiple frequencies with adequate angular resolution have often limited the use of these results for distinguishing clearly between the two scenarios, although recent results based on observations by NRH at metric wavelengths ($\sim$150--450 MHz) seem to favor Scenario 1 \citep{2012A&A...547A...6Z, 2014ApJ...795...68Z}. In addition, propagation effects of radio waves as they traverse the corona, which result in position shifts and angular broadening of the radio source, have further complicated the interpretation. For example, using observations by the LOw-Frequency ARray (LOFAR) near 30 MHz, \citet{2018ApJ...868...79C} reported a significant ($\sim$0.2 $R_{\odot}$) spatial separation of the HF and LF split-band lanes of an interplanetary type II burst , which, as the author argued, could be attributed to the scattering of radio waves from originally co-spatial sources. 

Here, first, thanks to our observations at high frequencies, the propagation effects are relatively insignificant (see, e.g., \citealt{1994ApJ...426..774B}). This is clearly demonstrated by the coherent dynamic evolution of the observed termination shock front (outlined by the spike source centroids) in response to the arrival of the plasma downflows \citep{2015Sci...350.1238C}. Second, our high positional accuracy ($<$1 Mm, or $<$20\% of the size of the shock front), dense frequency sampling, and ultra-high time cadence allow us to directly delineate the instantaneous morphology of the shock front for both the HF  and LF split-band lanes. Therefore, it is straightforward to explore the spatial relation between the HF and LF sources. In Figure \ref{fig:sbd_surface}(B), there are occasions when the HF and LF sources are nearly indistinguishable from each other within uncertainties (e.g., 18:30:09 UT). There are also other times when the two sources are separated along (i.e, in the horizontal direction, designated as $x$ hereafter; e.g., 18:30:17 UT) or across the shock front (in the vertical direction, designated as $y$ hereafter; e.g., 18:29:33 UT). As discussed earlier, the observed spatial distribution of both the HF and LF sources along the shock front clearly suggests that radio bursts of different frequencies (or plasma density) can be emitted from different sections of the shock front. Therefore, the termination shock is nonuniform along the shock front, which is by no means surprising given the dynamic and turbulent nature of the looptop region. We consider this nonuniformity as one piece of evidence that supports Scenario 2.

\begin{figure*}[ht!]
\epsscale{0.85}
\plotone{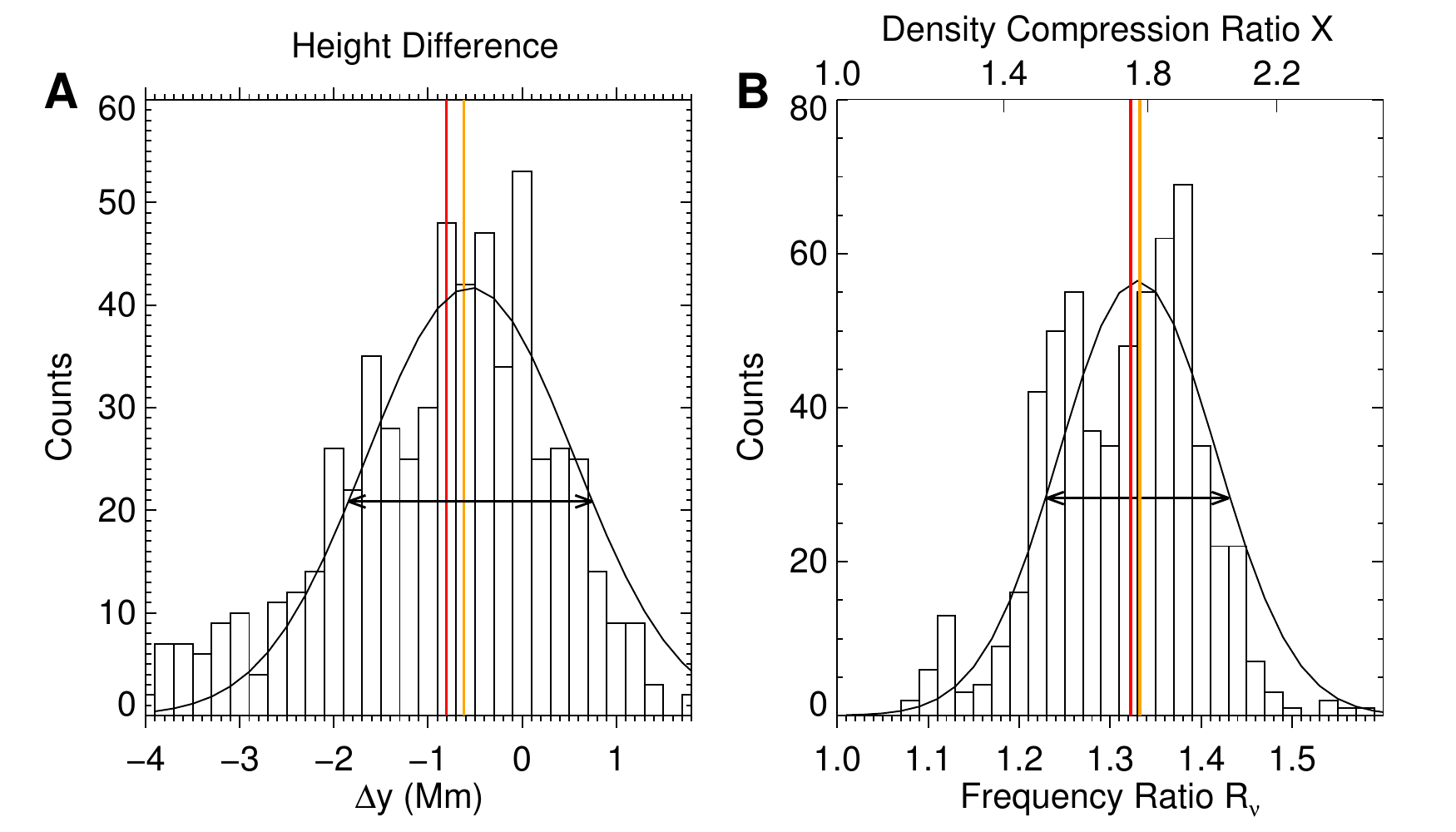}
\caption{Histogram of the height difference $\Delta y(x,t)$ (panel (A)) and frequency ratio $R_\nu(x,t)$ (panel (B)) between the HF and LF split-band sources at all times and locations along the shock. The $\Delta y$ distribution is skewed toward negative values, conforming with the expectation that the HF source is located in the shock downstream \textit{below} its LF counterpart. The corresponding density compression ratio $X(x,t)=R_\nu^2(x,t)$ is also shown in (B) as the top x-axis. Red and orange lines in both panels correspond to the average and median values of the distributions, respectively. Double-sided arrow in each panel indicates the full width half maximum (FWHM) of the distribution. \label{fig:hist}}
\end{figure*}

However, can we attribute the observed spatial separation between the HF and LF source purely to the nonuniformity along the shock front? An interesting impression from Figure \ref{fig:sbd_surface}(B) is that, for most times when both the HF and LF sources are seen at the same $x$ position along the shock front, the HF sources seem to be slightly \textit{below} the LF source. To better demonstrate this phenomenon, at each time $t$, we divide the shock front into multiple consecutive 1 Mm wide sections along the horizontal $x$-axis in Figure \ref{fig:sbd_surface}. For each shock section centered at a given $x$ position, we compute the height difference between the HF and LF source $\Delta y(x,t)=y_{\rm HF}(x,t)-y_{\rm LF}(x,t)$. By repeating such practice for all times and $x$ locations, we construct a distribution of $\Delta y(x, t)$. A histogram of $\Delta y(x, t)$ is shown in Figure \ref{fig:hist}(A). It is evident that the $\Delta y(x,t)$ distribution is heavily skewed toward negative values. The average value of $\Delta y$ is $\overline{\Delta y}=-0.80\pm0.02$ Mm (the quoted uncertainty here is the statistical standard error of the mean estimated as $\sigma/\sqrt{N}$, where $\sigma\approx 1.1$ Mm is the standard deviation of the distribution and $N$ is the sample size). This means that, at any given moment and at the same location of the shock front, the HF source is located \textit{immediately below} the LF source in a persistent manner. After separating the impacts of nonuniformity along the shock front and the shock geometry, here we provide clear evidence that supports the shock-upstream--downstream scenario for the LF and HF split-band sources (Scenario 1). We note that the particular shape of the $\Delta y(x,t)$ distribution may be intimately related to where the radio sources are emitted in the shock upstream and downstream plasma. However, it may also bear some contributions from the intrinsic uncertainty of our centroid locations ($<$1 Mm). Projection effects would also play a role, particularly if the shock front is not exactly viewed edge on. 

\subsection{Spatially and Temporally Resolved Shock Compression Ratio} \label{sec:compression}
In the previous subsection, we have provided strong evidence that supports the shock-upstream--downstream scenario to account for the observed LF and HF split-band feature of the termination shock event. Under this scenario, following the same technique for obtaining $\Delta y(x,t)$, at each time $t$ and spatial location along the shock front $x$, we can also compute the spatially and temporally resolved frequency ratio $R_\nu(x,t)=\nu_{\rm HF}(x,t)/\nu_{\rm LF}(x,t)$ and the associated density compression ratio $X(x,t)=R_\nu^2(x,t)$ along the shock front. A histogram for the distribution of $R_\nu(x,t)$ and $X(x,t)$ is shown in Figure \ref{fig:hist}(B). The values of $R_\nu$ and $X$ are distributed between 1.23--1.43 and 1.51--2.04, respectively (the ranges correspond to the FWHM of the respective distributions). The average values are $\overline{R_\nu}\approx1.33$ and $\overline{X}\approx1.78$. These numbers are broadly consistent with the frequency ratios reported in previous studies on metric and decametric type II radio burst events with split-band features \citep{2001A&A...377..321V,2012A&A...547A...6Z,2015ApJ...812...52D}. 

\begin{figure*}[ht!]
\epsscale{1.0}
\plotone{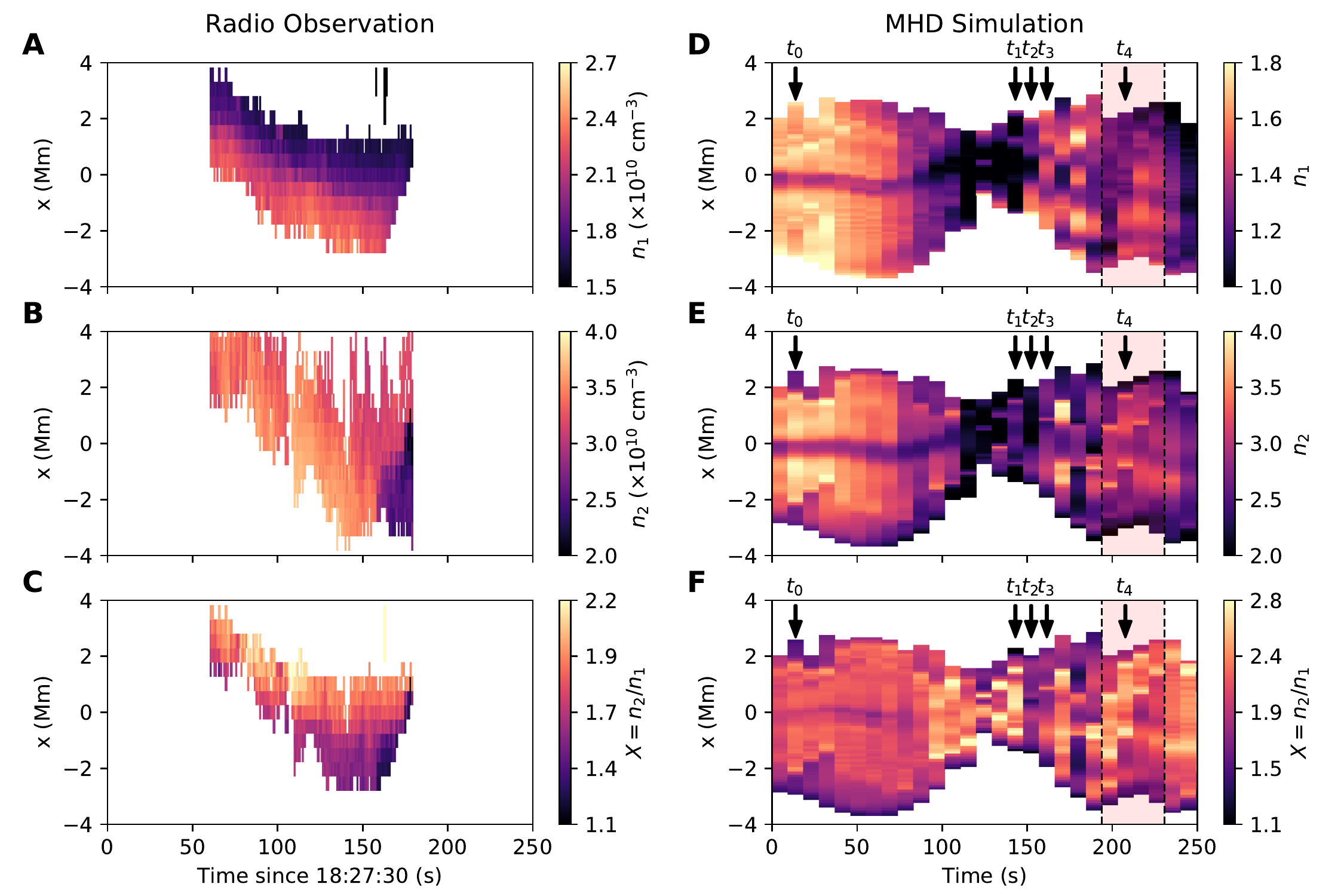}
\caption{(Left) Spatial and temporal variation of the upstream density $n_1$ (A), downstream density $n_2$ (B), and density compression ratio $X=n_2/n_1$ (C) derived from radio observation of the split-band feature. The horizontal axis is time, and the vertical axis corresponds to the spatial dimension along the shock front ($x$-axis in Figure \ref{fig:sbd_surface}). (Right) Same as the left panels, but showing all corresponding parameters from 2.5-D resistive MHD modeling. Labeled times indicate those selected in Figure \ref{fig:plasmoid}. The shaded time period (around $t_4$) indicates the times when a similar variation in density (and compression ratio) along the shock front is found in MHD modeling results. \label{fig:compression}}
\end{figure*}

The Mach number of MHD shocks can be inferred from the density compression ratio $X$ based on the Rankine-Hugoniot jump conditions. Yet this relation is a parametric function that include various plasma parameters in the upstream region and the angle between the shock normal and the magnetic field $\theta$, some of which are not constrained by our data. We refer to other works (e.g., \citealt{1995A&A...295..775M,2002A&A...396..673V,2014masu.book.....P}) for detailed discussions on the shock jump conditions in more general cases. Here, we only reiterate a few limiting cases in which the forms of the relation are greatly simplified. For perpendicular shocks (i.e., the angle between the magnetic field and shock normal $\theta_{Bn}=90^{\circ}$), the Alfv\'en Mach number $M_{\rm A}=v/v_{\rm A}$ (where $v_{\rm A}=B/\sqrt{4\pi n m_H}$ is the Alfv\'en speed) is related to the density compression ratio $X$ as
\begin{equation}
    M_{\rm A}=\sqrt{\frac{X(X+5+5\beta)}{2(4-X)}},
\end{equation}
where $\beta=8\pi nkT/B^2$ is the plasma beta (plasma-to-magnetic pressure ratio). In the low beta case ($\beta\to 0$), the relation simplifies to 
\begin{equation}
    M_{\rm A}=\sqrt{\frac{X(X+5)}{2(4-X)}}. \label{eq:lbeta}
\end{equation}
In the case of high plasma beta $\beta\to\infty$ or parallel shocks ($\theta=0^{\circ}$), the magnetic field plays a relatively limited role in the jump conditions other than changing the shock geometry, and the relation returns approximately to those of the hydrodynamic shocks:
\begin{equation}
    M_{\rm s}\approx\sqrt{\frac{3X}{4-X}}, \label{eq:hbeta}
\end{equation}
where $M_{\rm s}=v/c_{\rm s}$ is the sound-wave Mach number ($c_{\rm s}=\sqrt{\gamma kT/m_{\rm H}}$ is the sound speed of plasma with temperature $T$). For both the limiting cases in Eqs. \ref{eq:lbeta} and \ref{eq:hbeta}, the average value of the inferred Mach number is around 1.6, and the maximum can reach 2.0. Our measured density compression ratio and inferred Mach number of this termination shock case is similar to those predicted in the numerical modeling results of termination shocks \citep{1986ApJ...305..553F, 2011PhPl...18i2902W, 2018ApJ...869..116S}. They are also comparable to those estimated from split-band features in coronal-shock-driven type II radio bursts (see, e.g., statistical studies in \citealt{2002A&A...396..673V,2015ApJ...812...52D} and case studies in \citealt{2009ApJ...691L.151L,2012A&A...547A...6Z,2014ApJ...795...68Z,2018A&A...615A..89Z,2016ApJ...832...59K}). 

\begin{figure*}[ht!]
\epsscale{1.1}
\plotone{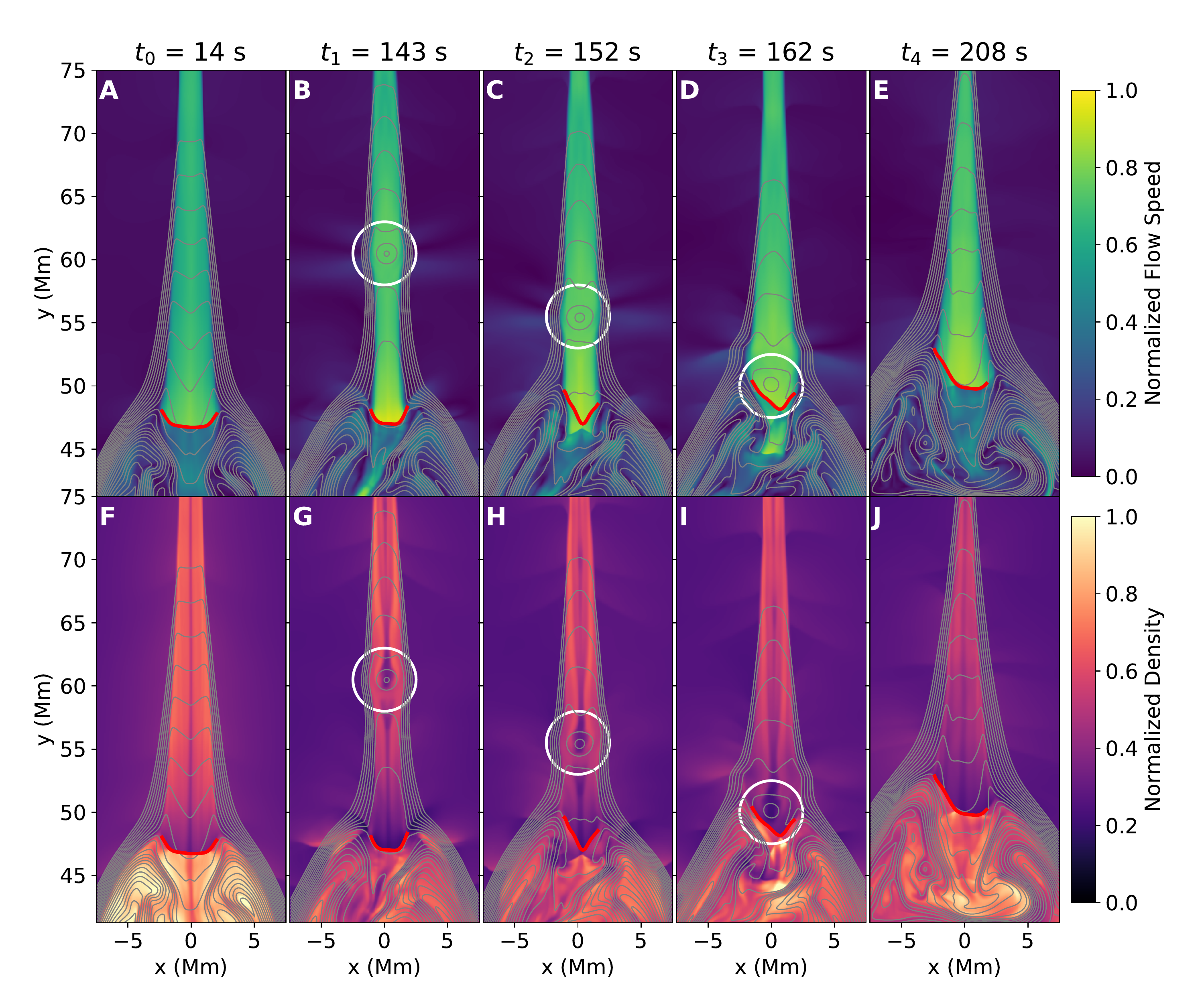}
\caption{MHD modeling of the reconnection downflow and the flare termination shock region at five selected times indicated in Figure \ref{fig:compression}(D)--(F). Colored background in the first and second row shows the flow speed (panels (A)--(E)) and plasma density (panels (F)--(J)) respectively. Magnetic field lines are shown as gray curves. The location of the termination shock in each frame is indicated by a thick red curve. The location of the plasmoid that causes distortion of the termination shock front is marked by a white circle. It propagates at a speed of $\sim$550 km s$^{-1}$. }\label{fig:plasmoid}
\end{figure*}

Previous theoretical studies have suggested that low-Mach-number, quasi-perpendicular flare termination shocks are capable of accelerating both electrons and ions efficiently, particularly when turbulence is present in the vicinity of the shock front to help the particles travel back and forth across the shock front to gain energy repeatedly \citep{2012ApJ...753...28G}. As shown by \citet{2015Sci...350.1238C}, the existence of density fluctuations in the vicinity of the shock front is strongly suggested by the presence of myriad short-lived stochastic spike bursts with a narrow frequency bandwidth, from which a density fluctuation level of $\delta n_e/n_e\approx 4\%$ was inferred. Their observations have further indicated that the flare termination shock should be capable of accelerating electrons to at least tens of keV. Moreover, if the flare termination shock operates in the high beta regime (defined as $\beta\gtrsim 1$), which is likely the case in the reconnection outflow region \citep{2013ApJ...766...39M, 2016ApJ...819...56S}, the flare termination shock may fall well into the supercritical regime for efficiently reflecting particles toward the upstream region, a condition favorable for ion acceleration (a shock is called supercritical if the downstream flow speed in the direction of the shock normal exceeds the magnetosonic speed; the critical value of $M_{\rm A}$ lies between 1.1 and 1.7 for quasi-perpendicular shocks in the high beta regime. See, e.g., \citealt{1984JPlPh..32..429E} and a recent review by \citealt{2009A&ARv..17..409T}).

More interestingly, some intriguing features are revealed in the spatially and temporally resolved ``time-distance'' maps of the upstream and downstream plasma density ($n_1(x,t)$ and $n_2(x,t)$), as well as the inferred density compression ratio $X(x,t)=n_2(x,t)/n_1(x,t)$, shown, respectively, in Figure \ref{fig:compression}(A)--(C). First, there is a persistent density gradient along the shock front from $-x$ to $+x$. This trend is much more prominent at the upstream side of the shock (panel (A)) than the downstream side (panel (B)), which, in turn, gives rise to a similar gradient for the density compression ratio along the shock front (panel (C)).

In order to understand the spatial and temporal variation of the observed density compression features along the termination shock front, we compare our observational results to 2.5D (resolved in $x$ and $y$, and uniform in the third $z$ dimension into the plane) resistive MHD modeling of the reconnection outflows and flare arcades with a Kopp--Pneuman-type reconnection geometry \citep{1976SoPh...50...85K}. In the MHD model, the reconnection current sheet has an edge-on viewing perspective similar to the event under study (i.e., current density $j$ in the current sheet is mostly along the $z$ direction). The simulation setup and main results on the termination shock dynamics were already presented in our previous works \citep{2015Sci...350.1238C, 2018ApJ...869..116S}, and we refer interested readers to \citet{2018ApJ...869..116S} for more detailed discussions. We outline the location of the termination shock front in the MHD model based on the divergence of the flow velocity $\nabla\cdot v$ (see, e.g., red curves in Figure \ref{fig:plasmoid}), and derive physical parameters including density, pressure, temperature, magnetic field, and velocity at both the upstream and downstream side of the shock front, as well as the associated shock compression ratios and Mach numbers.

In Figures \ref{fig:compression}(D)--(F), we show the equivalent time--distance maps of upstream density $n_1$, downstream density $n_2$, and density compression ratio $X$ for a selected period of time in the MHD modeling results. At certain times (marked by the shaded region), $n_1$, $n_2$, and $X$ at the termination shock front share the similar behavior as the observations: there is an evident gradient of $n_1$ and $X$ along the shock front that decreases toward the $+x$ direction, while the downstream density $n_2$ shows relatively smaller change. At other times (e.g., earlier in the figure at time stamps of $\sim$0--70 s), however, the shock is more or less symmetric and has less profound spatiotemporal variations.

What is the cause for the asymmetry and spatial variations along the shock front in the MHD simulation? In the simulation, since we do not impose an explicit symmetry about $x=0$, small-scale fluctuations that naturally arise in the numerical simulation can cause slight asymmetries about the $x$ axis. Yet the overall reconnection geometry is largely unaffected by the fluctuations so long as the reconnection proceeds in a steady fashion. However, when plasmoids are formed, they grow rapidly due to the highly nonlinear nature of the plasmoid instability and develop into a variety of sizes and shapes \citep{2018ApJ...869..116S}. When these plasmoids impinge upon the termination shock, a distorted and asymmetric shock front is expected. This is demonstrated in Figure \ref{fig:plasmoid}: before the formation of a plasmoid, the reconnection outflows on the shock upstream side are relatively steady, and the shock front is nearly symmetric about the center (panel (A)). When the plasmoid forms and arrives at the termination shock, a distorted and asymmetric shock front is observed (panels (B)--(E)). 

\begin{figure}[ht!]
\epsscale{1.1}
\plotone{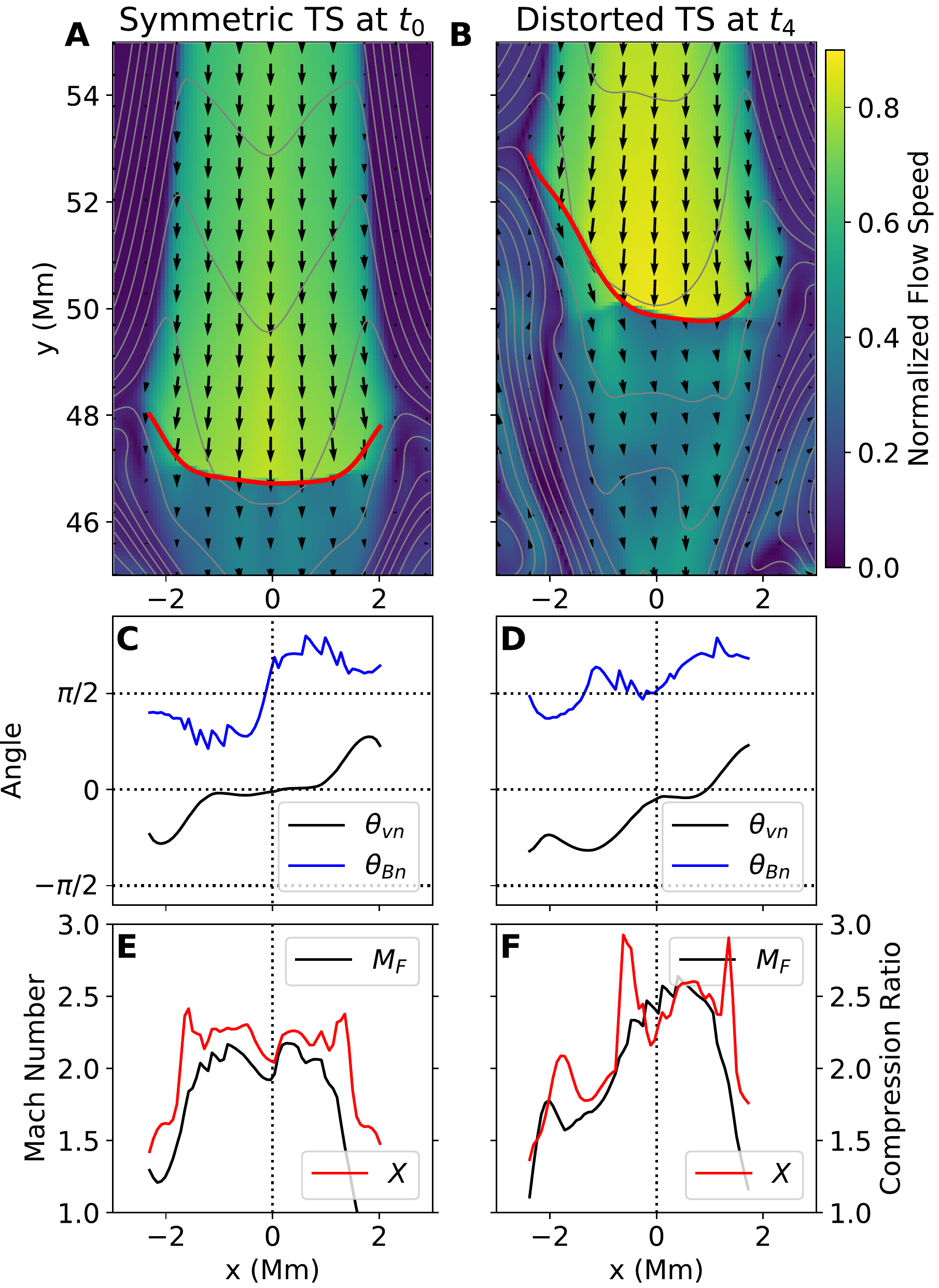}
\caption{Shock angle, Mach number, and density compression ratio along the termination shock front for the symmetric (left column) and distorted (right column) case. (A) and (B) Detailed view of the termination shock region at $t_0$ and $t_4$ indicated in Figure \ref{fig:plasmoid}. Background color shows the flow speed, and the arrows indicate the flow field (whose lengths scale with the velocity magnitude). The shock front is marked by the thick red curve. Gray curves are magnetic field lines. (C) and (D) Variation of shock angles $\theta_{\rm vn}$ (angle between upstream flow velocity and shock normal; black curve) and $\theta_{\rm Bn}$ (angle between upstream magnetic field and shock normal; blue curve) along the shock front. (E) and (F) Variation of fast-mode magnetosonic Mach number $M_F$ (black curve) and density compression ratio $X$ (red curve) along the shock front. \label{fig:ts_surface}}
\end{figure}

In Figure \ref{fig:ts_surface}, we show a more detailed view of the shock morphology before and after the impact of the plasmoid (left and right panels correspond to $t_0$ and $t_4$ in Figure \ref{fig:plasmoid}, respectively). Before the plasmoid arrival when the shock front is nearly symmetric, the direction of the reconnection downflows nearly aligns with the shock normal (i.e., $\theta_{\rm vn}\approx 0$) for the central portion of the shock, with a pair of oblique shocks located at the flank (black curve in Figure \ref{fig:ts_surface}(C); similar to the results in \citealt{2015ApJ...805..135T}). After the arrival of the plasmoid, the left portion of the shock front ($-x$) is highly distorted, displaying a large angle against the direction of the reconnection outflows ($\theta_{\rm vn}\approx 60^{\circ}$). In both cases, the angle between the magnetic field and the shock normal $\theta_{\rm Bn}$ stays close to $90^{\circ}$, indicating that the termination shock is, in general, a quasi-perpendicular shock. 

\begin{figure*}[ht!]
\epsscale{1.2}
\plotone{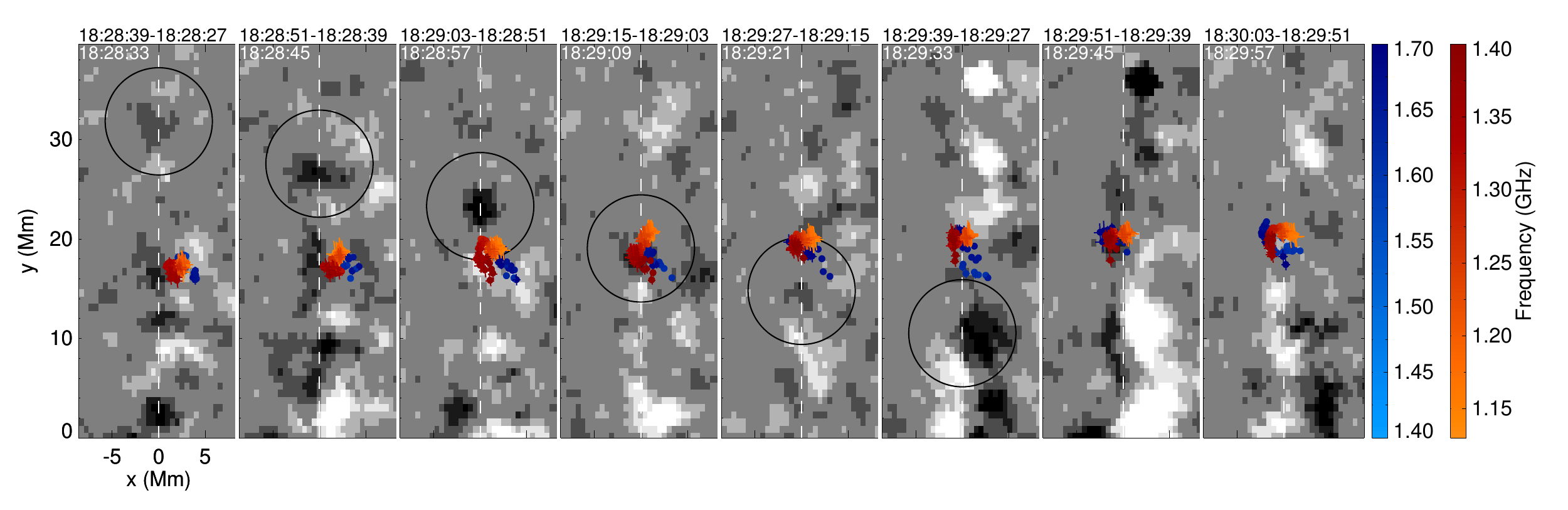}
\caption{SDO/AIA 94 \AA\ running difference images (grayscale background) showing a fast plasma blob impinging upon the termination shock front (black circles) when the split-band feature is present. The corresponding speed is $\sim$360 km s$^{-1}$ in projection. Color symbols are centriod locations of the split-band feature at different frequencies (same as Figure \ref{fig:sbd_surface}.) \label{fig:plasmoid_obs}}
\end{figure*}

Such a distorted shock geometry after the plasmoid arrival causes significant variations of the effective upstream flow speed in the direction normal to the shock surface $v'_{\rm n}=v_{\rm n}-v_{\rm n}^{\rm TS}$ (where $v_{\rm n}$ and $v_{\rm n}^{\rm TS}$ are the normal component of the flow speed and shock front speed, respectively). This effect, in turn, imposes profound impacts on the variation of the fast-mode Mach number $M_F=v'_{\rm n}/c_F$ (where $c_{\rm F}=\left[\frac{1}{2}\left(c_{\rm s}^2+v_{\rm A}^2+\sqrt{(c_{\rm s}^2+v_{\rm A}^2)^2-4c_{\rm s}^2v_{\rm A}^2\cos^2\theta_{\rm Bn}}\right)\right]^{1/2}$ is the fast-mode magnetosonic speed in the shock upstream) and the shock compression ratio $X$ along the shock front (shown in Figure \ref{fig:ts_surface}(E) and (F) as black and red curves, respectively). In particular, $M_{\rm F}$ and $X$ are strongly suppressed on the left ($-x$) side of the shock, largely due to the reduction of the effective upstream reconnection outflow speed $v'_{\rm n}$ into the shock front at a large incident angle $\theta_{\rm vn}$. 

In our observations, multitudes of fast plasma blobs are also seen in the EUV imaging data. They coincide with the distortion and disturbance of the shock front as outlined by the radio centroids of the stochastic spike bursts. Some of the plasma blobs cause a nearly total destruction of the shock front (\citealt{2015Sci...350.1238C}; occurred about one minute later than the split-band feature presented here), which coincides with a significant reduction of nonthermal radio and X-ray flux at both the looptop and loop-leg locations. As interpreted in \citet{2015Sci...350.1238C}, such a correlation serves as a strong evidence that supports the flare termination shock as a driver for accelerating nonthermal electrons to at least tens of keV. In the present case, however, the shock surface (as outlined by the centroids of the stochastic spike bursts shown in Figure \ref{fig:plasmoid_obs}) remains in place but only gets distorted in its location and morphology. In our MHD modeling, both cases---partial distortion and total destruction of the shock front---can be found. These dynamics are largely determined by the detailed properties of the plasmoids, most notably their size and momentum. We direct interested readers to \citet{2018ApJ...869..116S} for more detailed discussions. 

The observed plasma blob impinging upon the termination shock front during the split-band feature has a speed of $\sim$360 km s$^{-1}$ in projection, which is relatively slow comparing to the simulation ($\sim$550 km s$^{-1}$ for the plasmoid shown in Figure \ref{fig:plasmoid}) and other times when the stochastic spike bursts are present ($\sim$550 km s$^{-1}$ on average; \citealt{2015Sci...350.1238C}). Nonetheless the speed is comparable to the local sound speed $c_s\approx 329$--466 km s$^{-1}$ for 5--10 MK plasma (indicated by the presence of AIA 94 and 131 \AA\ emission). We note that, however, there is overwhelming evidence, both in observations and theoretical/numerical studies, which suggests that the speeds of the observable moving features in the plasma sheet (sometimes referred to as ``supra-arcade downflows,'' ``plasmoids,'' or ``contracting loops'' in the literature) tend to be considerably slower than the presumably Alfv\'{e}nic reconnection outflows \citep{1999ApJ...519L..93M,2004ApJ...605L..77A,2011ApJ...730...98S,2011ApJ...737...14S,2013ApJ...767..168L,2018ApJ...868..148L}. Understanding the seemly sub-Alfv\'{e}nic plasma flow features is a subject of ongoing research. One possibility involves plasma flowing within a high-$\beta$ plasma sheet that has a low Alfv\'{e}n speed. Another explanation argues that these features may be plasma structures embedded in the current sheet, which are not necessarily the Alfv\'{e}nic reconnection outflows themselves (see, e.g., discussions in \citealt{2018ApJ...868..148L}). In any case, we argue that it is highly probable that the reconnection outflows can well exceed the observed speeds of the plasma downflows, and can be supermagnetosonic to drive a fast-mode flare termination shock with a Mach number up to 2.0.

\begin{figure*}[ht!]
\epsscale{1.0}
\plotone{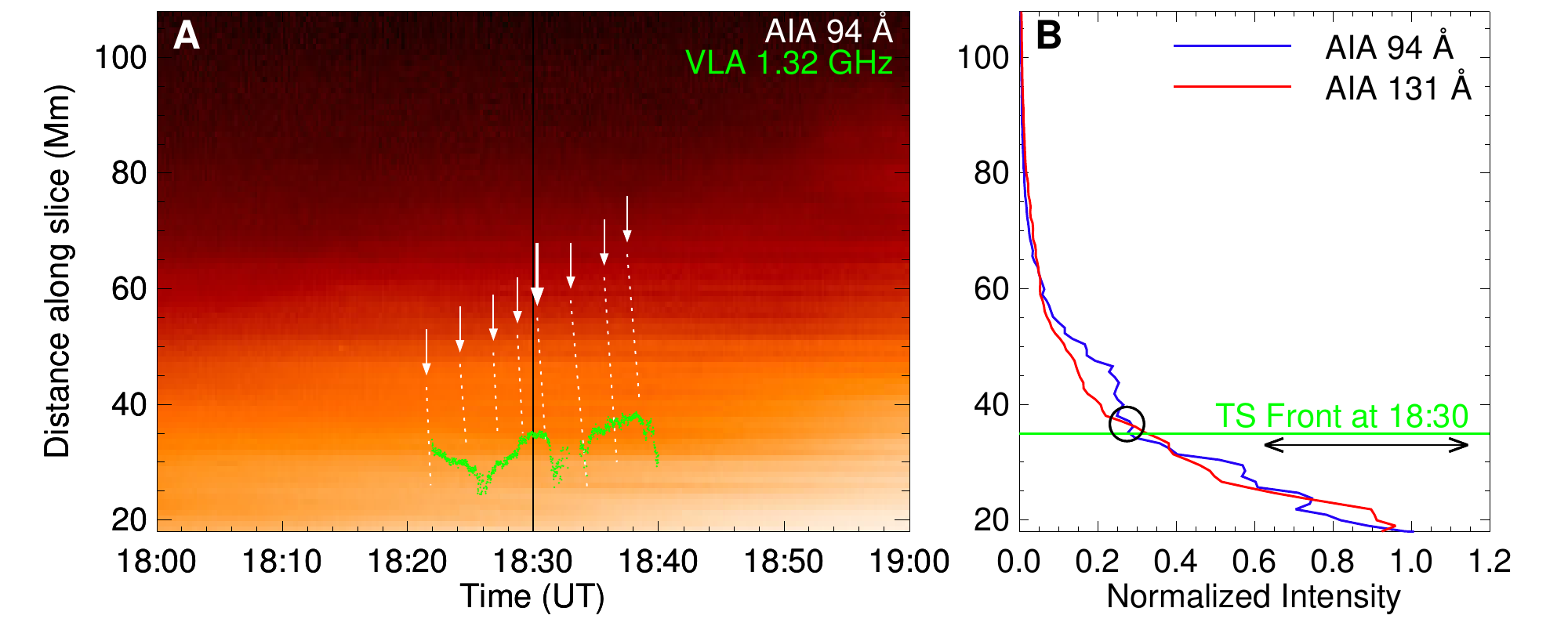}
\caption{EUV intensity variation across the termination shock front. (A) Detailed view of the \textit{SDO}/AIA 94 \AA\ time--distance plot in the looptop region obtained along the dotted line in Figure \ref{fig:overview}. The location of the radio spike sources are shown as green symbols. Some plasma downflows are marked as dotted lines (revealed in running-ratio plots; see Figure 2(C) in \citealt{2015Sci...350.1238C}). (B) AIA 94 \AA\ (blue) and 131 \AA\ (red) intensity variation along the slice at 18:30 UT (black vertical line in (A)). The location of the stochastic spike bursts is indicated by the green horizontal line. The double-sided arrow shows the shock upstream intensity (marked as a black circle) scaled up by a factor of $X^2$. \label{fig:stackplot}}
\end{figure*}

The close similarities between the observations and the simulations on the variations in upstream/downstream density and compression ratio along the shock front, as well as the presence of fast plasma blobs, strongly support that the observed split-band features are due to emission from both the upstream and downstream side of a highly dynamic termination shock. We would like to caution that, however, the observations are essentially 2D projections of a dynamic three-dimensional shock structure; therefore, the projection effects could play a certain role in the interpretation of the observational results. In addition, our 2.5D MHD simulation has its own limitations: not only does it lack the third dimension, but also, for example, the Lundquist number $S$ (a dimensionless ratio between the Alfv\'{e}n wave crossing time to the timescale of resistive diffusion), which has a strong impact on the production of the plasmoids, is orders of magnitude lower than the realistic values (which is the case for virtually all present MHD simulations). Hence our simulation does not necessarily reproduce every detail of the shock, including, for instance, the detailed spatial gradient and timescale of the shock compression. Therefore, our observation-simulation comparison described here should only be regarded as a qualitative demonstration, but not as a comprehensive reproduction of all of the observed features of the termination shock.

\subsection{Lack of Density Compression Signature in EUV} \label{sec:euv}

Since we have seen a significant density compression of $X\approx1.5$--2.0 across the shock front and a density variation of up to $\sim$80\% and along the shock front, an interesting question arises: can we observe the same termination shock feature in high-resolution \textit{SDO}/AIA EUV imaging data? The EUV intensity in AIA passband $i$ at pixel ($x$, $y$) is given by
\begin{equation}
I_i(x,y)=\int_{T}\left[G_i(T)d\xi_C(x,y,T)/dT\right]dT,\label{eq:euv}
\end{equation}
where $G_i(T)$ is a temperature dependent response function of this passband, and $d\xi_C(x,y,T)/dT \approx d(n^2\Delta h)/dT$ is the differential emission measure (DEM) along the line of sight (LOS) with a column depth of $\Delta h$ \citep[e.g.,][]{2010A&A...521A..21O,2012A&A...539A.146H}. If all other conditions are equal, we have $I \propto n^2$. In this case, $I$ is projected to show similar (and in fact, more profound, due to the quadratic dependence) spatial-temporal variation features to $n$ in the vicinity of the termination shock as shown in Figures \ref{fig:compression}(A)--(C). Moreover, $I$ would display a sharp increase across the termination shock front by a factor of $X^2$. Both signatures, however, are absent from the EUV imaging data. First, as shown in Figure \ref{fig:schem}(B), while the density/frequency decreases toward the right ($+x$) direction along the termination shock front (see also Figure \ref{fig:compression}(A)), the background intensity in AIA 94 \AA\ shows an evident increase, which runs opposite to the $I\propto n^2$ dependence. Second, no sharp EUV intensity jump (by a factor of $X^2$) is found across the shock front: in Figure \ref{fig:stackplot}(A) and (B), the EUV intensity, made at a slice along the direction of the cusp loops (dashed line in Figure \ref{fig:overview}), only displays a rather smooth increase toward lower heights. Moreover, the EUV intensity at the immediate downstream of the termination shock $I_2$ is much lower than the intensity expected from the shock compression, which is the upstream intensity $I_1$ scaled up by a factor of $X^2$ (indicated by the range bracketed by a double-sided arrow in Figure \ref{fig:stackplot}(B)).

We suspect that the LOS column depth ($\Delta h$) may play an important role in such an apparent discrepancy: while the plasma density in the close vicinity of the termination shock (derived from the radio emission frequency) is independent from the column depth, the EUV intensity $I$, however, has the contribution from all the plasma along LOS, which include but are not limited to the plasma at the termination shock. In fact, due to the rather stringent condition for the formation of the termination shock (i.e., $v'_{n}>c_F$), the shock front may likely only occupy a small column depth along the LOS direction. In this case, the ``shocked'' plasma may contribute to an insignificant portion of all of the emission measure along the LOS ($\xi_C$). Therefore, the shock density compression and its variations are effectively ``buried'' in the observed EUV intensity. 

Another possibility for the lack of an observable sharp contrast in EUV intensity may be due to the viewing geometry. If the shock front deviates slightly from the edge-on perspective (inferred from the flare geometry in Figure \ref{fig:overview}), the EUV intensity in the immediate vicinity of the shock front would have the contribution from both the upstream and downstream plasma along the LOS. In this case, the intensity jump across the shock front is effectively ``smeared out'' and the expected sharp contrast becomes more difficult to observe. We note that such a potential deviation from the edge-on perspective has little impact on the radio split-band feature in the frequency-time domain, but may contribute to broadening the distribution of the relative height between the HF and LF source $\Delta y$ (as shown in Figure \ref{fig:hist}(A)).

Finally, the limited angular resolution of \textit{SDO}/AIA and its filter-band-based imaging technique may also play a certain role in the non-detection of a sharp intensity jump in EUV. First, as inferred from the radio spectroscopic imaging data, the density jump across the shock front occurs only at a spatial scale of $\sim$0.8 Mm, which is close to or smaller than AIA's resolution limit (1$''$.5 or 1.1 Mm; \citealt{2012SoPh..275...17L}). Second, from Eq. \ref{eq:euv}, the EUV intensity at each filter band $I_i$ is also a function of the temperature response function $G_i(T)$. Hence the resulting intensity contrast has a strong dependence on the temperature variation across the shock front, which is expected from shock jump conditions. DEM reconstruction based on EUV intensity from multiple bands may give us some insights on this temperature variation. However, it is beyond the scope of the current study to investigate this aspect in more detail, particularly in light of the multiple challenges discussed above.

Similar phenomena of radio sources that lack corresponding EUV/X-ray signatures in solar flares have been frequently reported in the literature \citep{2013ApJ...763L..21C,2018ApJ...866...62C,2017ApJ...845..135F,2018ApJ...852...32K}. One of the most notable is \dml\ type III radio bursts: despite that recent dynamic spectroscopic imaging with high centroiding accuracy allows to precisely map detailed trajectories of type-III-burst-emitting electron beams propagating along newly reconnected magnetic flux tubes with arcsecond-scale accuracy, no corresponding loop-like structures have been found in high-resolution \textit{SDO}/AIA EUV imaging data \citep{2013ApJ...763L..21C,2018ApJ...866...62C}. Another type of such phenomena are microwave gyrosynchrotron sources (emitted by nonthermal electrons of hundreds of keV) that sometimes lack observable counterparts in EUV and/or X-rays, which are often shown to occupy large, tenuous coronal loops \citep{2017ApJ...845..135F,2018ApJ...852...32K}. Interpretation for such ``EUV/X-ray-invisible'' radio sources has been largely along the same line of argument: the LOS column depth or emission measure associated with the radio-emitting plasma is insufficient for the source to be observed as distinguishable features against the background by current EUV/X-ray instrumentation. Our study, once again, demonstrates the unique power of radio bursts in detecting and diagnosing certain key phenomena in solar flares and the need for future EUV/X-ray facilities with better angular resolution, temporal cadence, and temperature diagnostics.

\section{Conclusion}\label{sec:conclusion}
We report a detailed study of the split-band feature present in the coherent radio bursts associated with the arguably best studied flare termination shock event to date by \citet{2015Sci...350.1238C}. The split-band feature, a phenomenon well-known in type II radio bursts associated with fast-propagating coronal shocks, appears as two lanes in the radio dynamic spectrum separated in frequency, each of which consist of myriad short-duration, narrow-bandwidth stochastic spike bursts. By using high-cadence dynamic spectroscopic imaging with precise centroid locating capabilities ($<$1.$''$3, or $\sim$1 Mm) offered by the Jansky VLA, we determine the respective locations of the two split-band lanes as a function of frequency and time. We find that, while both of the split-band lanes each delineate a surface-like feature, the high-frequency lane is located consistently below the low-frequency lane by $\sim$0.8 Mm, which strongly support the shock-upstream--downstream interpretation for split-band features: The high- and low-frequency lanes are emitted in the downstream and upstream side of the termination shock front, respectively, with their frequency ratio $R_\nu=\nu_{\rm HF}/\nu_{\rm LF}$ determined by the density compression ratio $X=n_2/n_1=R_{\nu}^2$ across the shock front. 

Under this scenario, we derive spatially and temporally resolved density compression ratio at the termination shock front. The average compression ratio is $\overline{X}\approx 1.78$ and the inferred shock Mach number is $\sim$1.6 on average, but can reach up to $\sim$2.0. The shock compression ratio and Mach number are consistent with earlier numerical simulation results, and are comparable to those inferred from split-band features in type II radio bursts. Although this flare termination shock is probably a low-Mach-number shock, observational evidence has shown that it is capable of accelerating nonthermal electrons to at least tens of keV.

The spatial variation of $X$ shows an asymmetric feature, which has a consistently higher value at the right ($+x$) side of the termination shock front. We compare the observed spatiotemporal variation features with state-of-the-art 2.5D resistive MHD modeling results performed by \citet{2018ApJ...869..116S}. We find that such an asymmetry in shock compression and Mach number share close similarities with the MHD results when the termination shock front becomes asymmetric due to the impacts of fast plasmoids. Similar fast plasma downflows are also observed in the EUV difference imaging data when the split-band feature is present. We conclude that the detailed variations of the shock compression and Mach number may be due to the impact of fast plasma structures that distort the dynamic shock front, although the projection effects along the line of sight can not be completely ruled out. 

\acknowledgments

The NRAO is a facility of the National Science Foundation (NSF) operated under cooperative agreement by Associated Universities, Inc. This work made use of software packages including CASA \citep{2007ASPC..376..127M}, SunPy \citep{2015CS&D....8a4009S}, Astropy \citep{2013A&A...558A..33A, 2018AJ....156..123A}, and Athena \citep{2008ApJS..178..137S}. B.C. and S.Y. are supported by NASA grant NNX17AB82G and NSF grants AGS-1654382, AGS-1723436, and AST-1735405 to New Jersey Institute of Technology. K.R. and C.S. are supported by NASA grant NNX17AB82G and NSF grants AGS-1723425, AGS-1723313 and AST-1735525 to Smithsonian Astrophysical Observatory. F.G. acknowledges the support from NSF grant AST-1735414 and DOE grant DE-SC0018240.

\vspace{5mm}
\facilities{VLA, \textit{SDO}, \textit{RHESSI}}

\software{CASA, SunPy, Astropy, Athena.}

\bibliography{ts_splitband}


\end{document}